\documentclass[11pt]{article}

\usepackage{graphicx}
\usepackage{bm}
\usepackage{epsf}
\usepackage{rotating}
\usepackage{epsfig,graphics,rotate,color}
\usepackage{wrapfig}  
\usepackage{amssymb}
\usepackage{amsmath}
\usepackage{amsfonts}
\usepackage{array,hhline,dcolumn}%

\advance \topmargin by -\headheight
\advance \topmargin by -\headsep      
     
\evensidemargin \oddsidemargin
\begin{document}

~~~~~

\begin{center}
\noindent {\bf {\Large Whitepaper on the DAE$\delta$ALUS Program}} \\
 {\it The DAE$\delta$ALUS Collaboration}, \today 
{\vspace{0.1in} \hrule}

\end{center}
\vskip 0.5cm

\begin{center}

{\it C. Aberle$^5$, A. Adelmann$^{15}$, J.Alonso$^{12}$, W.A. Barletta$^{12}$, R. Barlow$^9$,
L. Bartoszek$^3$,  A. Bungau$^9$, A. Calanna$^{12}$, D. Campo$^{12}$,
L. Calabretta$^{10}$, L. Celona$^{10}$,G. Collin$^{12}$,
 J.M. Conrad$^{12}$,\\ A. de Gouv\^ea$^{14}$, Z. Djurcic$^2$,  S. Gammino$^{10}$, D. Garisto$^7$, 
 R. Gutierrez$^5$, R.R. Johnson$^4$,\\
 Y. Kamyshkov$^{17}$, G. Karagiorgi$^7$,
A. Kolano$^9$, F. Labrecque$^4$, W. Loinaz$^1$, H. Okuno$^{16}$, \\ V. Papavassiliou$^{13}$, K. Scholberg$^8$, M.H. Shaevitz$^7$,
I. Shimizu$^{18}$, J. Spitz$^{12}$, M. Skuhersky$^5$,\\ K. Terao$^7$,
M. Toups$^{12}$, M. Vagins$^{5,11}$,
D. Winklehner$^{12}$,  L.A. Winslow$^5$, J.J. Yang$^{12}$}
\end{center}

\vskip 1.cm

\begin{center}
$^1$Amherst College, Amherst MA, U.S.\\
$^2$Argonne National Laboratory, Argonne IL, U.S.\\
$^3$Bartoszek Engineering, Aurora IL, U.S.\\
$^4$Best Cyclotron System, Inc., Vancouver BC, Canada\\
$^5$University of California, Los Angeles, CA, U.S.\\
$^6$University of California, Irvine, CA, U.S.\\
$^7$Columbia University, New York NY, U.S.\\
$^8$Duke University, Durham NC, U.S.\\
$^9$University of Huddersfield, Huddersfield, U.K.\\
$^{10}$INFN-LNS, Catania, Italy \\ 
$^{11}$Institute for the Physics and Mathematics of the Universe, Kashiwa, Japan\\
$^{12}$Massachusetts Institute of Technology, Cambridge MA, U.S.\\
$^{13}$New Mexico State University, Las Cruces NM, U.S.\\
$^{14}$Northwestern University, Evanston IL, U.S.\\
$^{15}$Paul Scherrer Institute, Villigen, Switzerland\\
$^{16}$RIKEN, Wako, Japan\\
$^{17}$University of Tennessee, Knoxville TN, U.S.\\
$^{18}$Tohoku University, Sendai, Japan\\

\end{center}

\vskip 0.5cm

\noindent {\it Abstract:}   This whitepaper describes the status of
the DAE$\delta$ALUS program for development of high power
cyclotrons as of the time of the final meeting of 
the Division of Particles and Fields 2013 Community Study
(``Snowmass'').     We report several new results, including a
measurement capability between $\sim$4 and 12 degrees on the $CP$ violating parameter
in the neutrino sector.     Past results, including the capability of 
the IsoDAR high $\Delta m^2$ $\bar \nu_e$ disappearance search, are
reviewed.    A discussion of the R\&D successes, including
construction of a beamline teststand,  and future plans are
provided.  This text incorporates short whitepapers written for
subgroups in the Intensity Frontier and Frontier Capabilities Working
Groups that are available on the Snowmass website.

\newpage

\pagestyle{plain}

\section{Introduction}

DAE$\delta$ALUS (Decay-At-rest Experiment for $\delta_{CP}$ studies
At the Laboratory for Underground Science) is a phased R\&D
program leading to a high-sensitivity search for $CP$-violation 
\cite{EOI, firstpaper}.     This is a unique, cyclotron-driven
$\bar \nu_\mu \rightarrow \bar \nu_e$ search for a non-zero
$CP$ violation parameter,
$\delta_{CP}$, in the three-neutrino mixing matrix.  
DAE$\delta$ALUS, when combined with Hyper K (with-JPARC beam), can 
achieve an uncertainty of 4 to 12 degrees on $\delta_{CP}$--well beyond the
sensitivity of LBNE or HyperK alone.   The system consists of a
two-cyclotron design.   The smaller injector cyclotron, which will be
developed first, also can be used as a driver to provide a very pure $\bar  \nu_e$ flux.
This can be paired with KamLAND to allow for 
a disappearance search, called IsoDAR,  that is a
factor of five times more sensitive to oscillations indicative of a sterile neurino
than other proposals.    DAE$\delta$ALUS and IsoDAR are two
examples of applications of these machines.    However, one can
envision uses beyond these within neutrino physics, including cross
sections and Beyond Standard Model Searches.    These cyclotrons are
also valuable commercially, which is why DAE$\delta$ALUS has a strong 
industry-university collaboration on the R\&D program.

This whitepaper reports on the status of the DAE$\delta$ALUS program 
for the Division of Particles and Fields 2013 Community Study
(``Snowmass'').   We provide an overview of the physics:  the $CP$ Violation
search,   the $\bar \nu_e$ disappearance search, the measurement of
cross sections; and the search for new physics.    We then provide
a discussion of status and plans:  an introduction to the context;  progress
on the design;  the studies at a Best Cyclotrons, Inc., teststand and the planned
Catania teststand;  and the required longer term R\&D.    We end by
discussing the broader impacts of these machines.  This
whitepaper incorporates the text from the many one- or two-page
whitepapers that the DAE$\delta$ALUS collaboration has provided to the
Neutrino Working Group (Intensity Frontier), the Working Group on High Intensity
Secondary Beams Driven by Protons (Frontier Capabilities) and the 
Working Group on Accelerator Technology testbeds and test Beams 
(Frontier Capabilities).    As we combine the content of these whitepapers, we
maintain the philosophy that each section should be short, but
well-justified through the references, which is possible because of
extensive documentation concerning the program already on the arXiv
and in published journals.

\section{Physics of The Program}

\subsection{DAE$\delta$ALUS:   The $CP$ Violation Search \label{CP}}

\begin{figure}[p]
{\includegraphics[width=6.in]{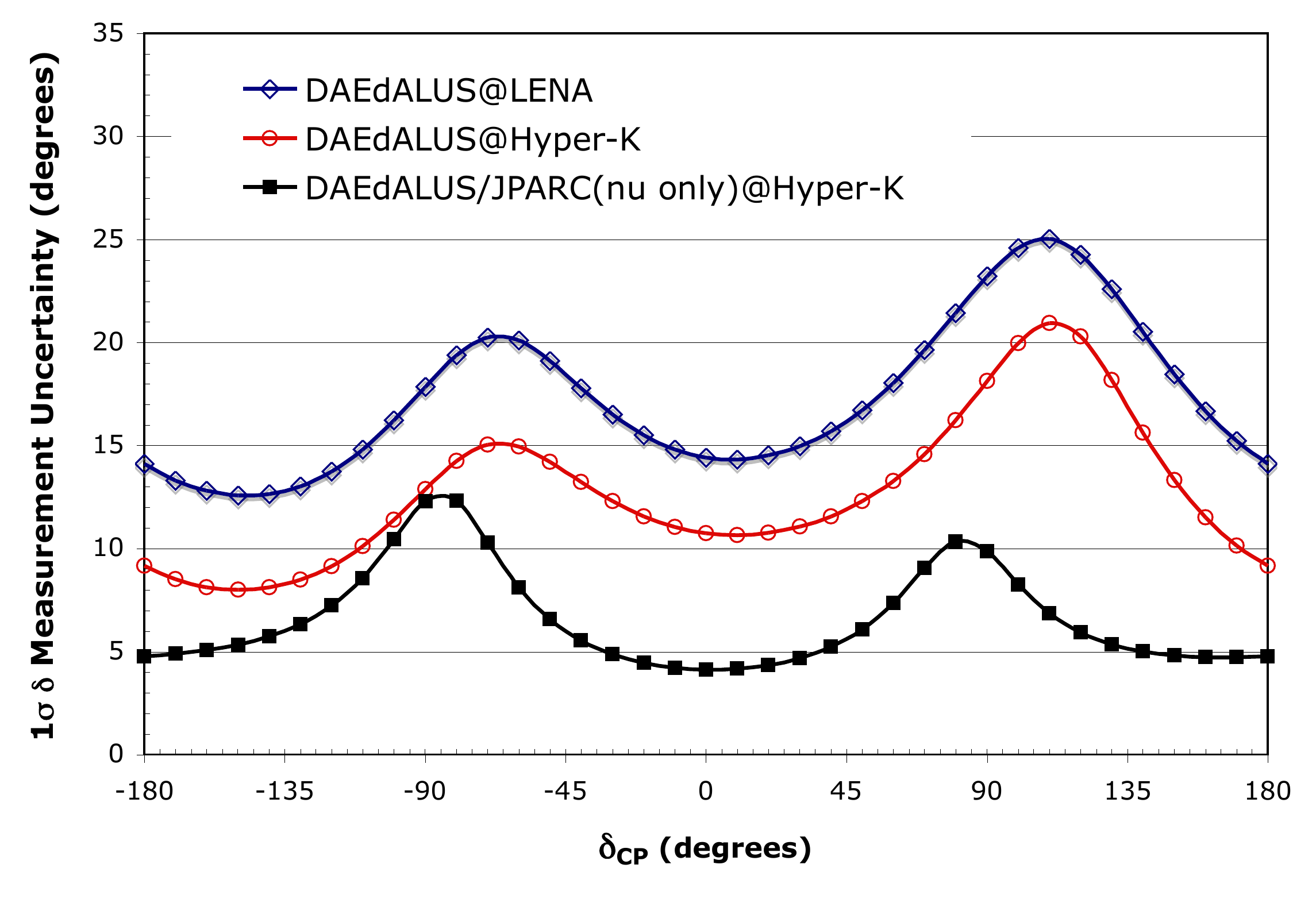}\hfill}
{\includegraphics[width=6.in]{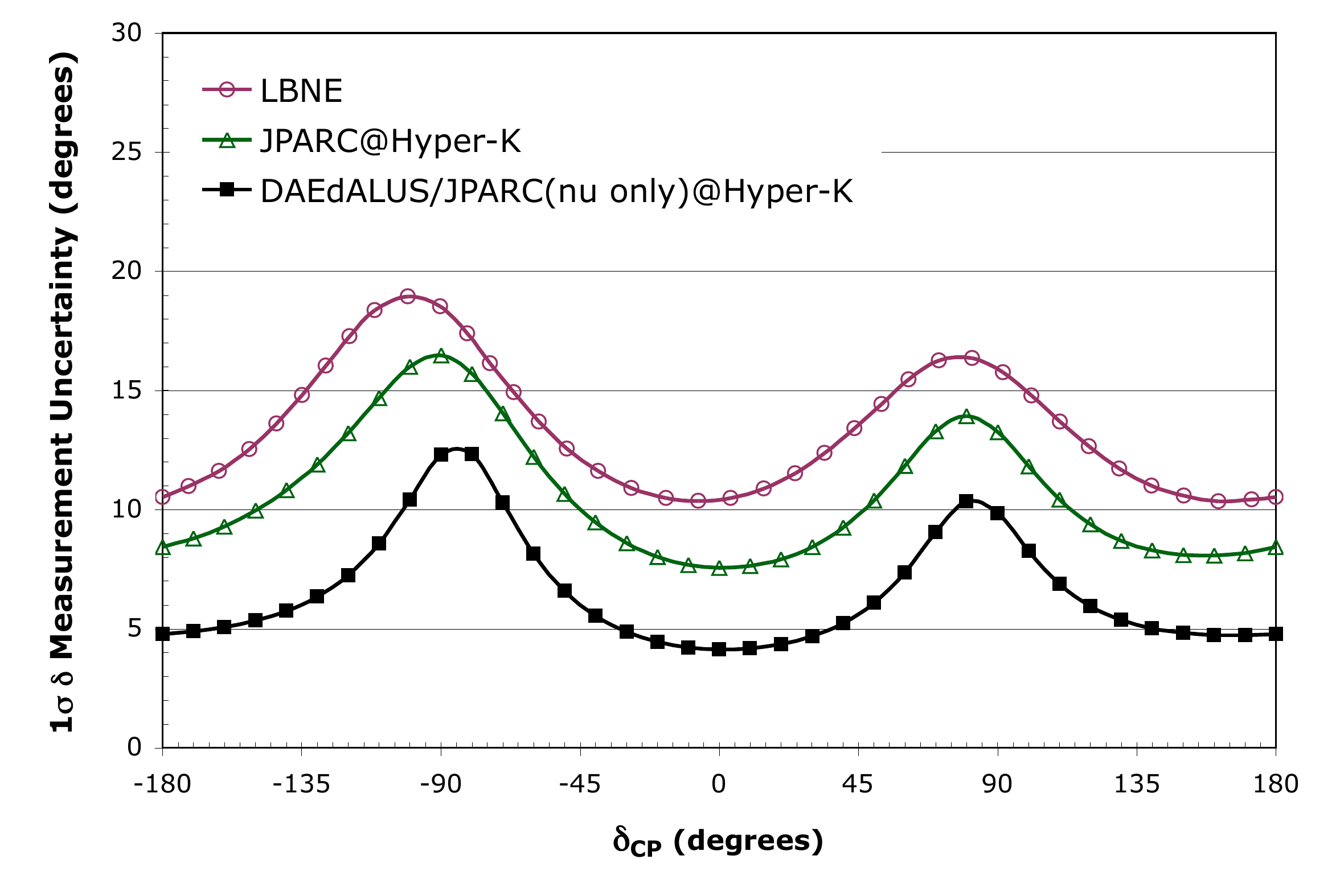}\hfill}
\caption{\it \footnotesize  Top:
The sensitivity of the $CP$-violation search in various
configurations:  Dark Blue -- DAE$\delta$ALUS@LENA,
Red-DAE$\delta$ALUS@Hyper-K,
Black--DAE$\delta$ALUS/JPARC(nu-only)@Hyper-K.
Bottom:   Light Blue-- LBNE;  Green-- JPARC@Hyper-K \cite{hyperk}
Black--DAE$\delta$ALUS/JPARC(nu-only)@Hyper-K
(same as above).   See Table~\ref{tab:configs} for description of each
configuration.    
\label{Dsense}}
\end{figure}

The DAE$\delta$ALUS phased
program for $CP$-violation studies is formed by sets of ``modules'' consisting
of  an ion source, an injector cyclotron (DIC), a superconducting
ring cyclotron (DSRC),
and a target/dump.    Modules are
arranged at three sites. 
The experiment uses $\pi$/$\mu$ decay-at-rest (DAR) to search for $\bar{\nu}_{\mu} \rightarrow \bar{\nu}_e$
oscillations with the atmospheric $\Delta m^2$. The $\bar{\nu}_e$ will be detected,
via the inverse beta decay interaction, in
an underground, ultra-large free-proton-based (water or oil)  detector.
The $CP$-violation signal is extracted by measuring the oscillation 
wave as a function of distance $L$ that the neutrinos have travelled,
with positions at 
$L=$ 1.5 km, 8 km, and 20 km.  
The final phase of the program
leads to a unique $CP$-violation
experiment.  

The program proceeds in four phases.
Phase I, which is well underway, involves development and testing of an ion
source and low-energy beam transport system, including design of the
inflection system that guides the beam into the cyclotron.   Phase
II establishes the injector cyclotron, discussed in Sec.~\ref{disapp}, below.
Phase III will result in a running 
DSRC and associated target/dump; i.e. the
first full accelerator module.  This module will be located
nearest to the detector and runs with a 13\% duty factor,
producing 0.8 MW average power on the dump/target.
The physics case for this module
is based on short-baseline searches for Beyond Standard Model 
physics \cite{WeakSanjib, ArgawallaConradShaevitz}.
Lastly,   Phase IV introduces 
the modifications for high-power running needed at the mid and far
sites for $CP$-violation studies.

\begin{table}[t]
\centering {\footnotesize
\begin{tabular}{|l|c|c|c|c|c|}
\hline
Configuration  &  Source(s) & Average& Detector & 
Fiducial & Run \\ 
Name  &   & Long Baseline &  &  
Volume & Length \\  
 &   & Beam Power &  &  
 &  \\ \hline  
DAE$\delta$ALUS@LENA & DAE$\delta$ALUS only &  N/A & LENA & 50 kt & 10
years \\   \hline
DAE$\delta$ALUS@Hyper-K & DAE$\delta$ALUS only &  N/A & Hyper-K & 560 kt & 10
years\\  \hline
DAE$\delta$ALUS/JPARC & DAE$\delta$ALUS  &   & Hyper-K & 560 kt & 10 
years\\ 
(nu only)@Hyper-K& \& JPARC  & 750 kW  &  & & \\  \hline
JPARC@Hyper-K & JPARC & 750 kW & Hyper-K & 560 kt & 3 years $\nu$ +\\
 & &  && & 7 
years $\bar \nu$ \cite{hyperk}\\  \hline
LBNE & FNAL & 850 kW & LBNE & 35 kt & 5 years $\nu$ \\
         &          &              &          &          &  5 years
         $\bar \nu$ \cite{LBNEPX}\\  
\hline
\end{tabular}}
\caption{\it \footnotesize  Configurations under study for
  Snowmass. \label{tab:configs} }
\end{table}

The program requires free proton targets, hence water or
scintillator detectors.
The original case was developed for
a 300 kt Gd doped water detector at Homestake, in
coordination with LBNE \cite{config}.   Subsequently,  DAE$\delta$ALUS was
incorporated into LENA \cite{LENA} (called ``DAE$\delta$ALUS@LENA'').   As a 50 kt scintillator oil
detector, LENA is substantially smaller than the original 300 kt water
design, but has the advantage of lower backgrounds.   The sensitivity of DAE$\delta$ALUS@LENA 
is shown in Fig.~\ref{Dsense} (Top).     For the Snowmass study, we 
have also considered a
phased program in Japan, beginning by pairing with the existing Super-K
detector (with Gd-doping) and followed
by running with 
a Gd-doped 560 kt Hyper-K \cite{hyperk} (``DAE$\delta$ALUS@Hyper-K'').  
This program could be combined with Hyper-K running with the 
750 kW JPARC beam (``DAE$\delta$ALUS/JPARC@Hyper-K'').  In this scenario, 
JPARC provides a pure $\nu_\mu$ flux,  which is the strength of a
conventional beam, while
DAE$\delta$ALUS provides a high statistics $\bar \nu_\mu$ flux.    This leads
to a impressive sensitivity to $\delta_{CP}$, as shown on Fig.~\ref{Dsense}~(Top).
A comparison with  the expectation of a 35 kt LBNE detector running at 850 kW
\cite{LBNEPX} and JPARC@Hyper-K \cite{hyperk} is shown in 
Fig.~\ref{Dsense}~(Bottom).   A summary of the assumptions for the
various
configurations is provided in Table~\ref{tab:configs}.
Further description of this study is in \cite{pontecorvo}.
A short-baseline beam from the ESS \cite{ESS} may also be  appealing,
as neither the DAE$\delta$ALUS nor ESS baselines are subject to matter
effects.   

This idea for a $CP$ violation search  has
been well received by
the wider community.  
The NRC Committee to Assess the Science
Proposed for a Deep Underground Science and Engineering Laboratory
wrote:
``Proposals for new second generation experiments with water
Cherenkov detectors include very imaginative possibilities, such as the
DAE$\delta$ALUS proposal to create neutrinos using a series of small nearby
cyclotrons'' \cite{NRC}.

\subsection{IsoDAR:  A Search for $\bar \nu_e$ Disappearance at Short
  Baseline \label{disapp}}

\begin{figure}[t]
\begin{center}
\vspace{-0.2in}
{\includegraphics[width=4.in]{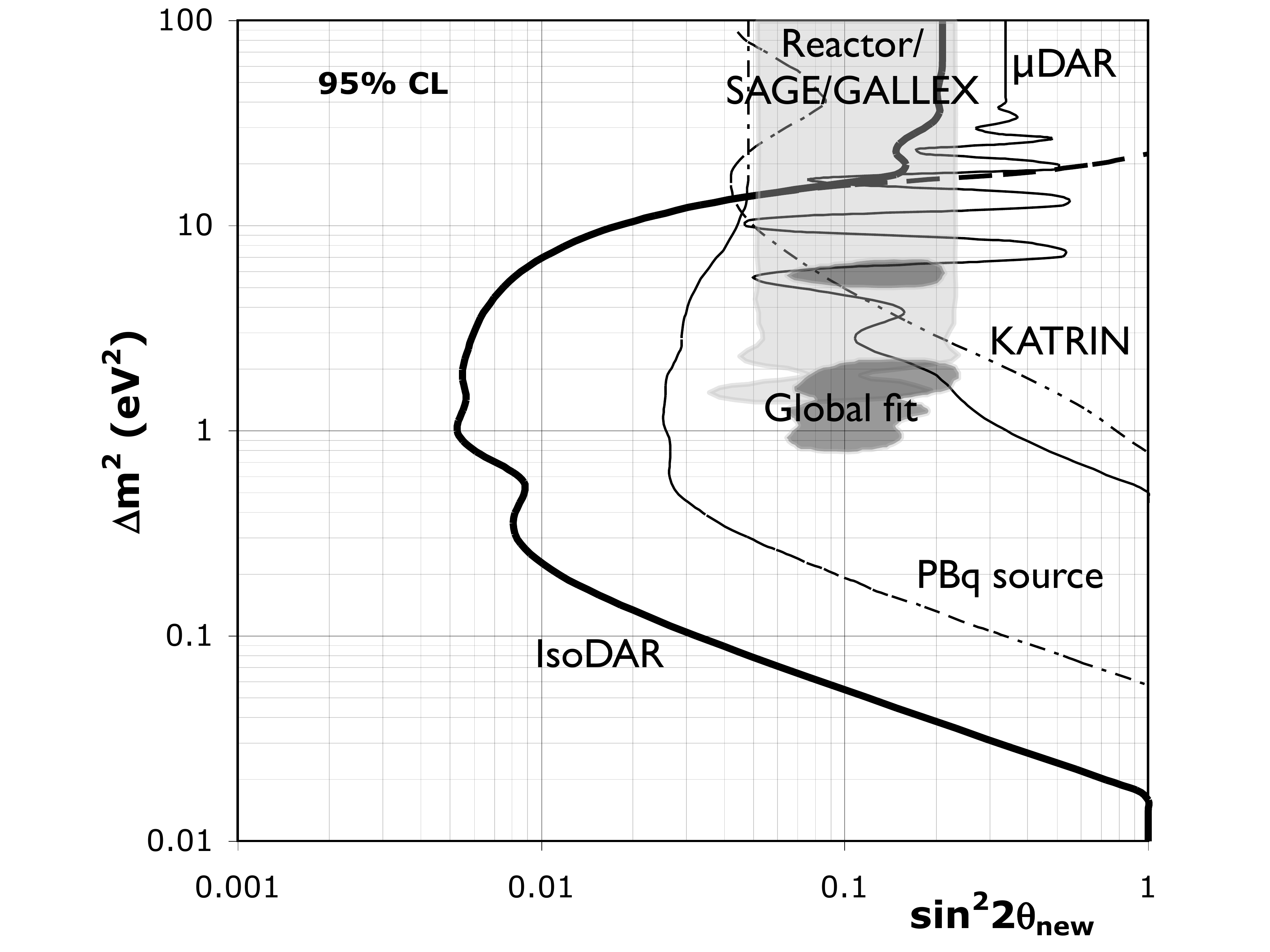}}
\vspace{-0.2in}
\end{center}
\caption{\it \footnotesize  
The sensitivity of the IsoDAR experiment to $\bar \nu_e$
disappearance in a five-year run, reprinted from Ref. \cite{prl}. The sensitivities for both
rate+energy shape (solid line) and shape-only (dashed line) are
shown.  Other curves  from Refs. \cite{ConradShaevitz, Mention,
Thierry, Formaggio}. 
\label{Isosensitivity}}
\begin{center}
{\includegraphics[height=2.in]{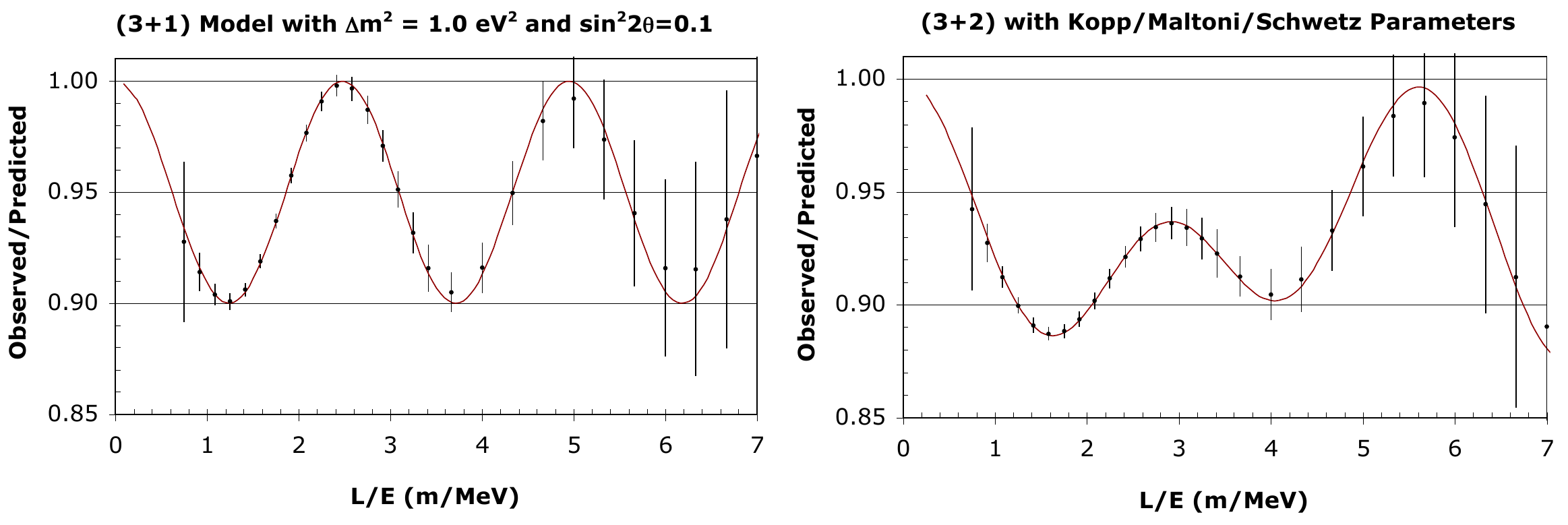}}
\vspace{-0.3in}
\end{center}
\caption{\it \footnotesize The $L/E$ dependence of two example oscillation
  signatures after five years of IsoDAR running considered in Ref.~\cite{prl}. 
The solid curve is the oscillation probability with no smearing in the
reconstructed position and energy. The 3+2 example (right) represents
oscillations with the global best fit 3+2 parameters from
Ref.~\cite{kopp}. 
\label{wiggles} }
\end{figure}

IsoDAR is a novel isotope decay-at-rest source of $\bar \nu_e$ for 
Beyond Standard Model searches.   
The source \cite{prl} consists of an accelerator
producing 60 MeV protons \cite{NIM} that impinge on a
$^9$Be target, producing neutrons.    IsoDAR can use the same cyclotron
design as the injector cyclotron for the two-cyclotron DAE$\delta$ALUS 
system.       The protons  enter 
a surrounding 99.99\% isotopically pure $^7$Li sleeve, where neutron
capture results in $^8$Li; this isotope undergoes
$\beta$ decay at rest to produce an isotropic
$\bar \nu_e$ flux with an average energy of $\sim$6.5 MeV and
an endpoint of $\sim$13 MeV.    
The $\bar \nu_e$ will interact in a scintillator detector via inverse beta decay (IBD),
$\bar \nu_e +p \rightarrow e^+ + n$, which is easily tagged through
prompt-light--neutron-capture coincidence.
When 
paired with KamLAND \cite{KLdet}, the experiment can observe
$8.2\times 10^5$ reconstructed IBD events in five years.
With this data set, IsoDAR will decisively test sterile neutrino
oscillation models, allow precision  measurement
of $\bar \nu_e$-$e$ 
scattering, and search for production and decay of exotic particles.

The sterile neutrino search uses
electron flavor disappearance, interpreted within models with three active
and one  (3+1) or more sterile neutrino flavors
\cite{Giunti1111.1069, kopp, sorel, 3+3}. 
These models result from fits to combined  appearance
(muon-to-electron flavor \cite{LSND, MBnu, MBnubar}) and
disappearance (muon flavor \cite{nomad, cdhs, ccfr84} and electron flavor
\cite{ConradShaevitz, Mention,
  sagegallex}) data. 
Fig.~\ref{Isosensitivity} presents the oscillation 
landscape at 95\% confidence level for electron flavor
disappearance in a
3+1 model.    IsoDAR covers the global fit allowed region for $\bar \nu_e$ disappearance
\cite{Giunti1111.1069}  (dark grey) at 
5$\sigma$ in four months.
The high statistics of the five-year run can
distinguish models with one or more sterile neutrinos, as shown in
Fig.~\ref{wiggles}.  The impressive capability of IsoDAR led to its
choice as a Physical Review Letters highlight \cite{highlight}.

\subsection{Precision Fluxes, Cross Sections and  Searches for New
  Physics \label{otherprec}}

The decay-at-rest (DAR) beams produced by the cyclotrons for the
DAE$\delta$ALUS and IsoDAR programs are unique sources for 
precision cross-section studies and high-sensitivity searches.   
As a result, interest has been expressed by the 
cross-section \cite{volpe,
  coherentdiscov, gail, strange} and 
exotic physics search communities \cite{jorge1, jorge2, kosto, cohosc,
  charged}.   Therefore, these cyclotrons can be seen as a
new tool within the overall neutrino program.   Stand-alone sites can
be implemented by coalitions of
universities within the neutrino
community.  
The relatively modest cost range (from \$25M to \$100M, depending on design) makes several centers potentially feasible.

\begin{wrapfigure}{l}{0.5\textwidth}
{\includegraphics[width=3.in]{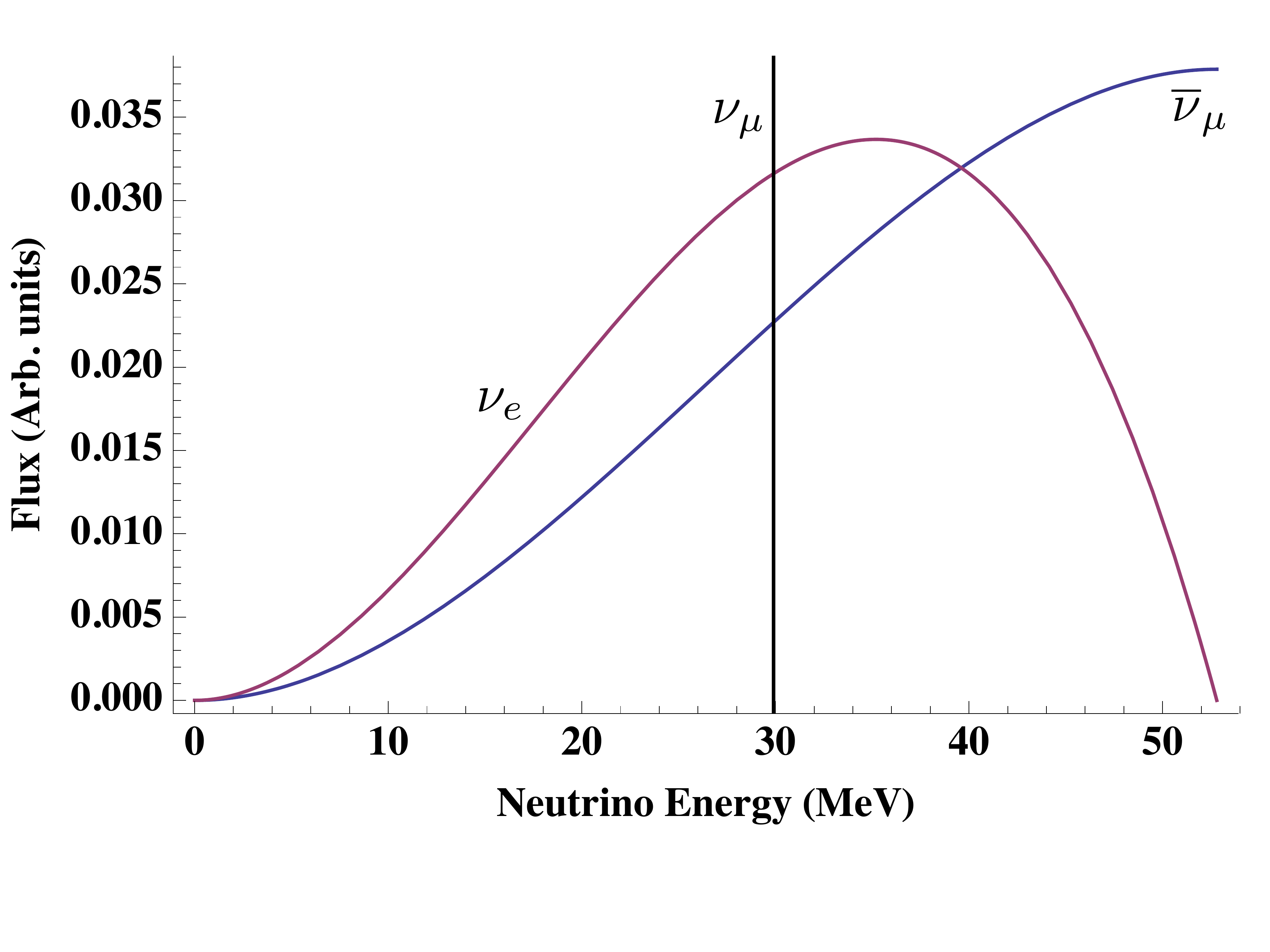}\hfill}\\
{\includegraphics[width=3.25in]{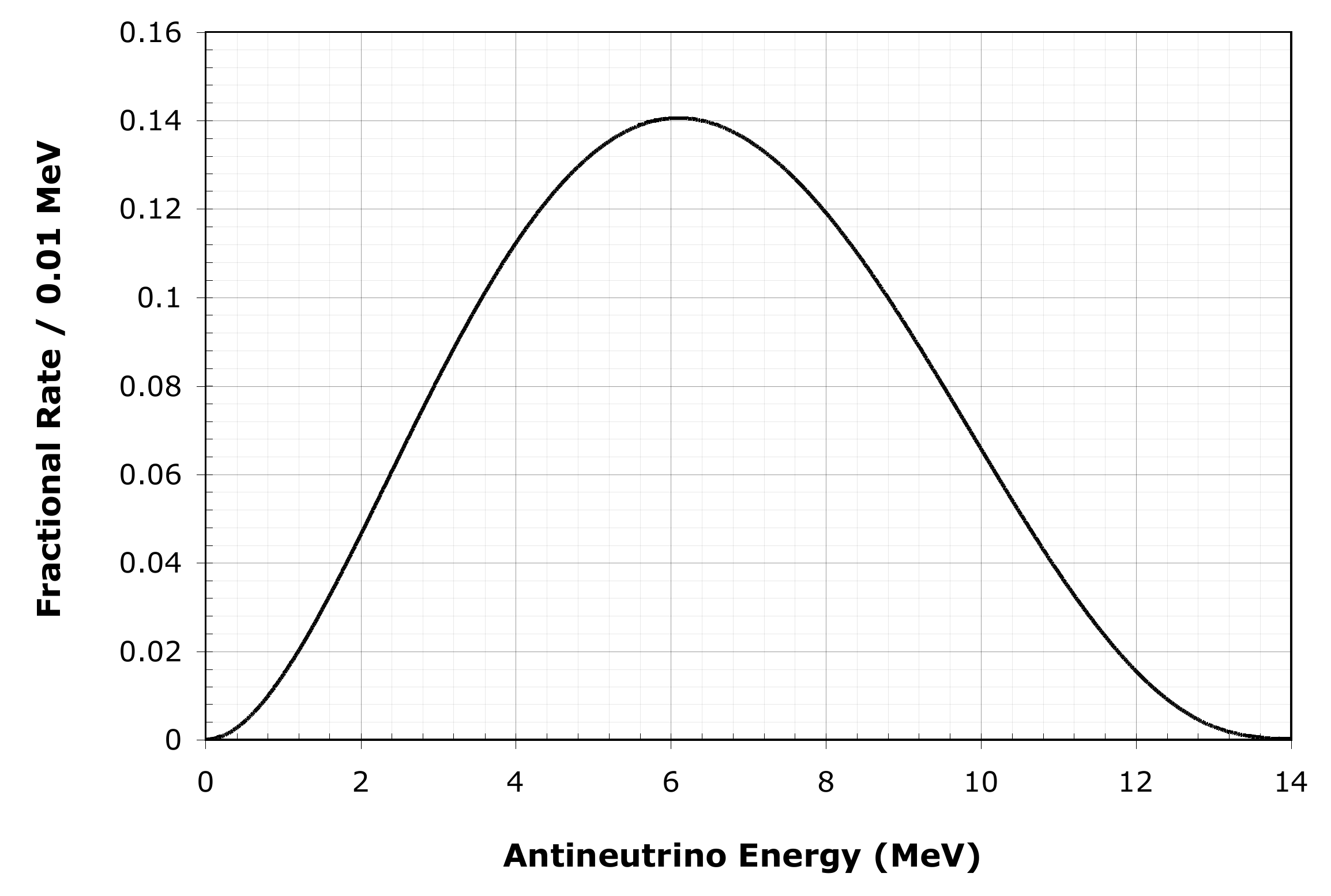}\hfill}
\caption{\it \footnotesize  
Fluxes from $\pi$/$\mu$ (top) and $^8$Li (bottom) decay at rest, from Ref.~\cite{pontecorvo}.
\label{fluxes}}
\end{wrapfigure}

As neutrino physics enters the precision era, $\pi$/$\mu$-DAR and
isotope-DAR beams (e.g., $^8$Li),
in which the energy dependence and the flavor contents are
precisely known (see Fig.~\ref{fluxes}), are valuable.   
There is already a history of cross-section physics from $\pi$/$\mu$
sources \cite{LSNDxsec, KARMEN, LAMPF}.  
The primary issue for precision measurement 
is that absolute flux rates are only 
predicted to $\sim$20\% \cite{burman}.    However,  one can potentially normalize to inverse
beta decay or to neutrino-electron scattering, both of which have cross
sections known to $<1\%$, greatly reducing normalization errors.

The sources can be designed to deliver $\sim$100 kW to 1 MW, depending 
on the flux required.  If space is limited, such as at an
underground lab, the optimal energy range for an $^8$Li DAR driver is about 60 
MeV/amu \cite{technology}.    
On the other hand, 
the driver energy is $\sim 800$ MeV/amu for
$\pi$/$\mu$ DAR sources\cite{EOI}.   
A variety of detectors can be paired with isotope and pion/muon DAR
sources, 
including dark matter detectors \cite{coherentdiscov} 
and scintillator detectors \cite{LSND, KARMEN}. 

Alternatively, the accelerators can provide charged
particle 
beams either directly (isotopes) or in a secondary line (pions, muons),
allowing applications beyond neutrino physics.   This has led to
worldwide interest.    For example, our 60 MeV/amu machine is quite
similar
to the cyclotron being developed for the
SPES (Selective Production of Exotic Species) Project \cite{SPES}, at
Laboratori Nazionali di Legnaro
in Italy.     Also,
our 800 MeV/amu cyclotron is quite similar to the existing
superconducting 
ring cyclotron of RIKEN used to drive their Radioactive Ion beam
facility \cite{Okuno}.
   
For the Snowmass study,  the collaboration is presenting searches for
nonstandard interactions (NSIs) through studies of antineutrino-electron scattering as an example of
a precision search and coherent
neutrino scattering as an example of cross-section studies.

\subsubsection{Antineutrino-Electron Scattering}

A sample of $> 2400$ antineutrino-electron scattering (ES) events ($\bar \nu_e
+ e^- \rightarrow \bar \nu_e + e^-$) will be collected by IsoDAR in a
5 year run--a
five times larger sample in the
low $Q^2$ range than exists today \cite{Irvine:1976, TEXONO:2012, ROVNO:1993, MUNU:2005}.
This allows for sensitive searches for
Beyond Standard Model physics through deviations of the cross section
from the {\it ab initio} prediction,  which is highly precise given
the recent electroweak results from LHC \cite{Erler}.    In our example,
we examine the effect of non-standard interactions on the measured weak couplings,  $g_R$
and $g_L$, or, equivalently, $\sin^2 \theta_W$.  
NSIs, introduced into the theory via an
effective 4-fermion term in the Lagrangian~\cite{Berezhiani:2001rs},
can induce instantaneous transitions from electron flavor to some
other flavor,  leading to deviations in the measurement from
expectation.   

The analysis method follows that of Ref.~\cite{ConradLinkShaevitz},
which proposed a similar measurement for reactors.  The IsoDAR-based
measurement has the advantage of a higher energy flux than that of a reactor.
The ES event rate is normalized by inverse beta decay events and the expected backgrounds for
a run at the KamLAND detector are well understood from
the solar neutrino studies \cite{kamsol} which are also a single-flash signal.

Our analysis will be described in Ref.~\cite{s2thw}, which is in
draft, and will appear on the arXiv in time for the final Snowmass
Meeting.   
Our study considered rate-only, energy-dependence-only, and
rate-plus-energy-dependence.   The strength of the result is in the
rate, with $\delta \sin^2\theta_W=0.0085$ for rate-only and 
$\delta \sin^2\theta_W=0.0084$ for rate-plus-energy-dependence.
Although backgrounds are constrained by beam-off running, subtraction
adds considerably to the error.    If directional reconstruction
becomes available through addition of fast PMTs that allow Cerenkov
reconstruction \cite{Lindley}, then the backgrounds may be reduced 
by a factor of two.  In this case, we achieve $\delta \sin^2\theta_W=0.0065$.

IsoDAR's sensitivity to the NSI  parameters $\epsilon_{ee}^{e
  L}$ and $\epsilon_{ee}^{e  R}$ are shown in 
Figure~\ref{precision} (left).    The IsoDAR result provides
complementary information to 
the currently allowed global regions 
for these parameters\cite{epsglob}.

\subsubsection{Coherent Neutrino and AntiNeutrino  Scattering}

Coherent neutrino scattering is an as-yet-unobserved Standard Model
process; however, its cross section is well-predicted.  This process
has a large cross section in the energy regime relevant for DAR
sources from the DAE$\delta$ALUS program.      However,  the
signature,  which is a recoil nucleon with $\sim 10$ keV, is quite
difficult to observe.  With this said, modern dark matter detectors
have this capability.   Thus, the injector cyclotron used in IsoDAR
or the full-scale DAE$\delta$ALUS system open up the opportunity
for discovering new physics through this precision measurement.

The collaboration has already published results on the sensitivity of
the $\pi/\mu$ DAR design to new physics through coherent scattering
\cite{coherentdiscov, cohosc}.    Therefore, here we focus on the latest
results which explore the expectation for the $^8$Li flux.  For
further details, see Ref.~\cite{pontecorvo}.

\begin{figure}[tb]
\begin{center}
\begin{tabular}{c c}
\includegraphics[scale=.32]{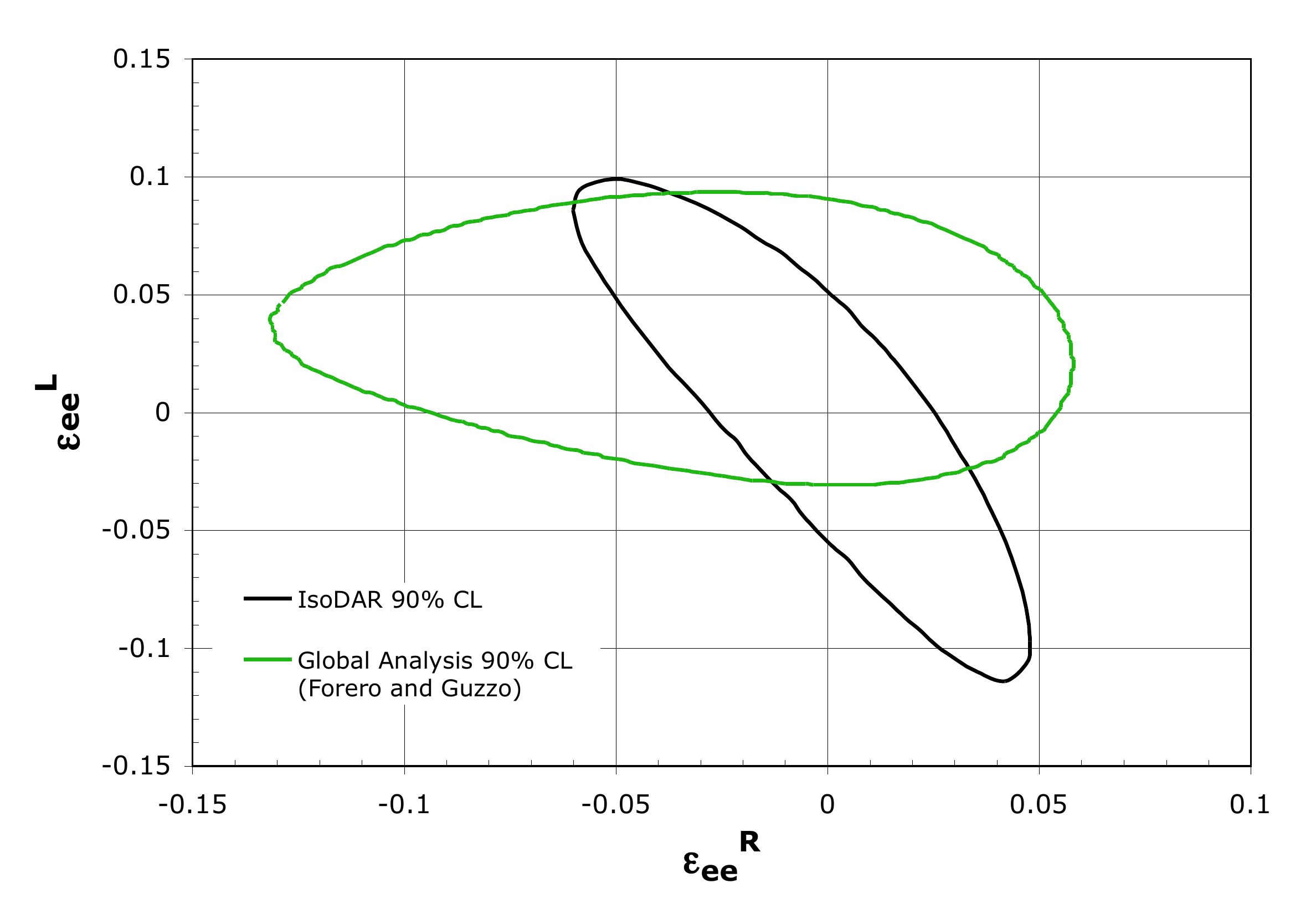}
&
{\includegraphics[scale=.45]{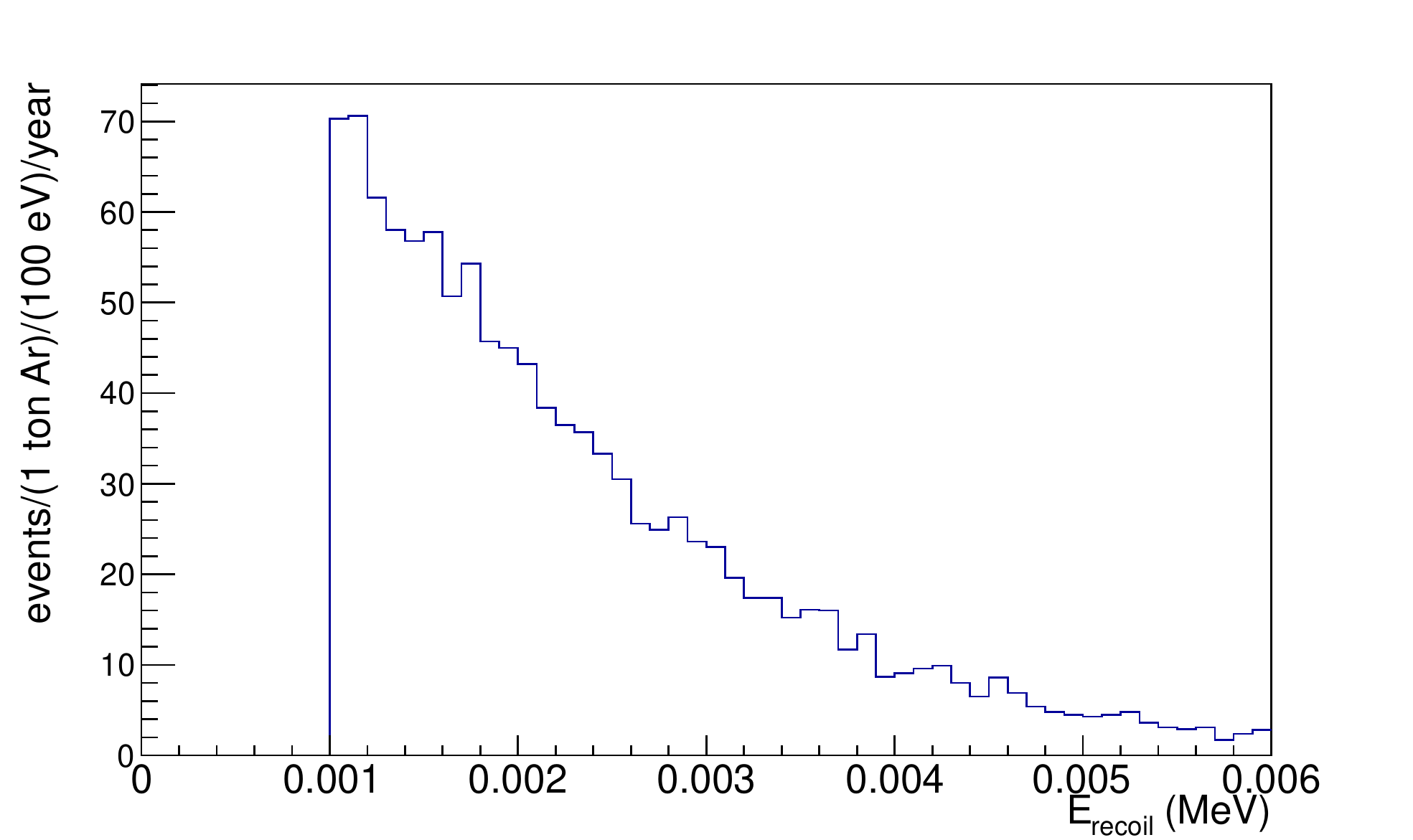}}
\end{tabular}
\end{center}
\vspace{-.5cm}
\caption{\it \footnotesize  Left: IsoDAR's 90\% CL sensitivity to $\epsilon_{ee}^{e
  L}$ and $\epsilon_{ee}^{e  R}$ compared to the present global
allowed region from Ref.~\cite{epsglob}.
          Further details will be provided in
          Ref.~\cite{s2thw}. Right: 
          Coherent event rate in terms of nuclear recoil energy with a
          1000~kg argon detector at a 10~m average baseline from the
          IsoDAR source..   Further explanation provided in Ref.~\cite{pontecorvo}.
\label{precision}}
\end{figure}

Figure~\ref{precision}(right)  shows the expected rates in terms of nuclear recoil energy for an IsoDAR source
(2.58$\times10^{22}~\overline{\nu}_e$/year) in combination with a 1000~kg argon detector at a
10~m average baseline from the source with a 1~keV nuclear recoil
energy threshold and 20\% energy resolution.  While such a detector is
not yet available,  this is within the goals of future dark matter
experiments.  Given these assumptions,
about 1200~events per year could be collected for a high statistics
sampling of this event class. A first observation of the process is
clearly possible with a more modest size detector as well.

\section{Accelerator Studies Status and Needs}

\subsection{Context for further discussion \label{gentleintro}}

The key to the program is state-of-the-art 
cyclotrons.   Before presenting specifics, it is worthwhile to briefly
review cyclotron design and terminology, since those reading this
report may not have experience with cyclotron accelerators.

In a ``compact'' cyclotron, charged particles
from an ion source are injected at the
center, often via a 
``spiral inflector'' -- a device that directs the beam from a vertical
(axial) direction
to the horizontal median plane of the cyclotron magnet.   Once inflected, the particles are 
bent by the magnetic field into roughly circular orbits. 
An RF cavity system accelerates the
particles and, as they gain energy, their trajectories spiral outwards.  At the outer radius
 a ``septum'' (thin sheet of conducting material such as carbon, copper, tantalum) placed
  between the Nth and (N+1)th turns enables bending the particles into an extraction channel
   by a strong electric field.  The spatial separation between the ``turns'' grows smaller
as the beam approaches the outer edge of the cyclotron, making extraction without losing 
a lot of beam on the septum more difficult for higher energies. 

A compact cyclotron uses a single coil surrounding the (generally very complicated)
 iron pole structure, and provides the smallest and most economical design for achieving
  the required beam energy.  The iron pole is generally machined with azimuthal
  ``sectors.''   The pole face will have ``hills'' and ``valleys''
  producing regions of high and low magnetic fields, which are needed to
  provide beam focusing and ``isochronicity.''  The isochronicity
  condition ensures the orbital revolution time is independent of beam 
  energy, so that the RF frequency does not need to change as the particles are accelerated.
  
  An alternate design, involving ``separated sector'' or individual 
  pie-shaped magnets, offers advantages for higher intensity beams at the cost of complexity,
   size, and increased capital costs.  The Paul Scherrer Institute (PSI) Injector II accelerator \cite{psicycloii}
   is of this type.  Beam can be injected into such a separated-sector
   machine at higher energies if one employs a 
    booster accelerator such as an RFQ or high-voltage platform.  This can mitigate
     much of the space-charge-induced reduction in beam quality of the compact, spiral-inflection design.  

Presently, the highest current achieved for axially-injected compact proton (or H$^-$) cyclotrons
 is $\lesssim$2 mA.  The separated-sector PSI injector runs at 3 mA.  To achieve the 
 required neutrino fluxes we need 10 mA of protons on target.  Our concept for achieving this 
 much higher current is to work with molecular hydrogen (H$_2^+$) beams.  As each ion carries two 
 protons, only 5 electrical milliamperes (emA) are required to produce
 10 mA of protons.  Furthermore, because of the doubled
  mass, the ions carry more momentum, and are thus less affected by the space-charge forces that tend to
   cause the beam size to increase.  These space-charge forces are characterized by a parameter 
   $K$ called ``generalized perveance.'' In fact, $K$ for a 5 emA H$_2^+$ at our planned injection energy of 
   70 keV (35 keV/amu) is similar to that of a 2 mA proton beam.  On this basis, we believe that we 
   can achieve the required currents on target using the less-expensive compact design.  Should it 
   be necessary, we can always fall back on the more expensive separated sector concept.

To achieve the 800 MeV energies needed for DAE$\delta$ALUS, the second
cyclotron (the DSRC) must be 
of the separated-sector design.  While the compact injector can have 
a normal-conducting (iron-dominated) magnet, the higher energy machine must have superconducting 
coils and quite high ($\sim$6T) peak fields.  Injection is achieved, with the 60 MeV/amu energy with a series 
of electrostatic and magnetic elements that nudge the beam into the lowest-radius orbit.  RF cavities 
again accelerate the beam to high energies, as quickly as possible to avoid beam losses due to 
stripping in the residual gas of the cyclotron.  Extraction is done by a stripping foil; the H$_2^+$ ion enters
 the foil but emerges as two protons.  These spiral inwards, but because of the highly irregular magnetic
  field a stripper-foil location can be found that enables the protons to exit the machine cleanly, about 180 
  degrees from the foil position.  This stripping technique avoids the need for clean turn separation, which 
  as mentioned earlier becomes very difficult at high energies.  The only consequence is that ions hitting the foil
   could come from one of several turns (hence have an 
   energy spread $\sim$1$\%$).  
   It is easy to design the extraction channel to have momentum acceptance adequate to accommodate this.

\begin{wrapfigure}{l}{0.5\textwidth}
{\includegraphics[width=3.5in]{{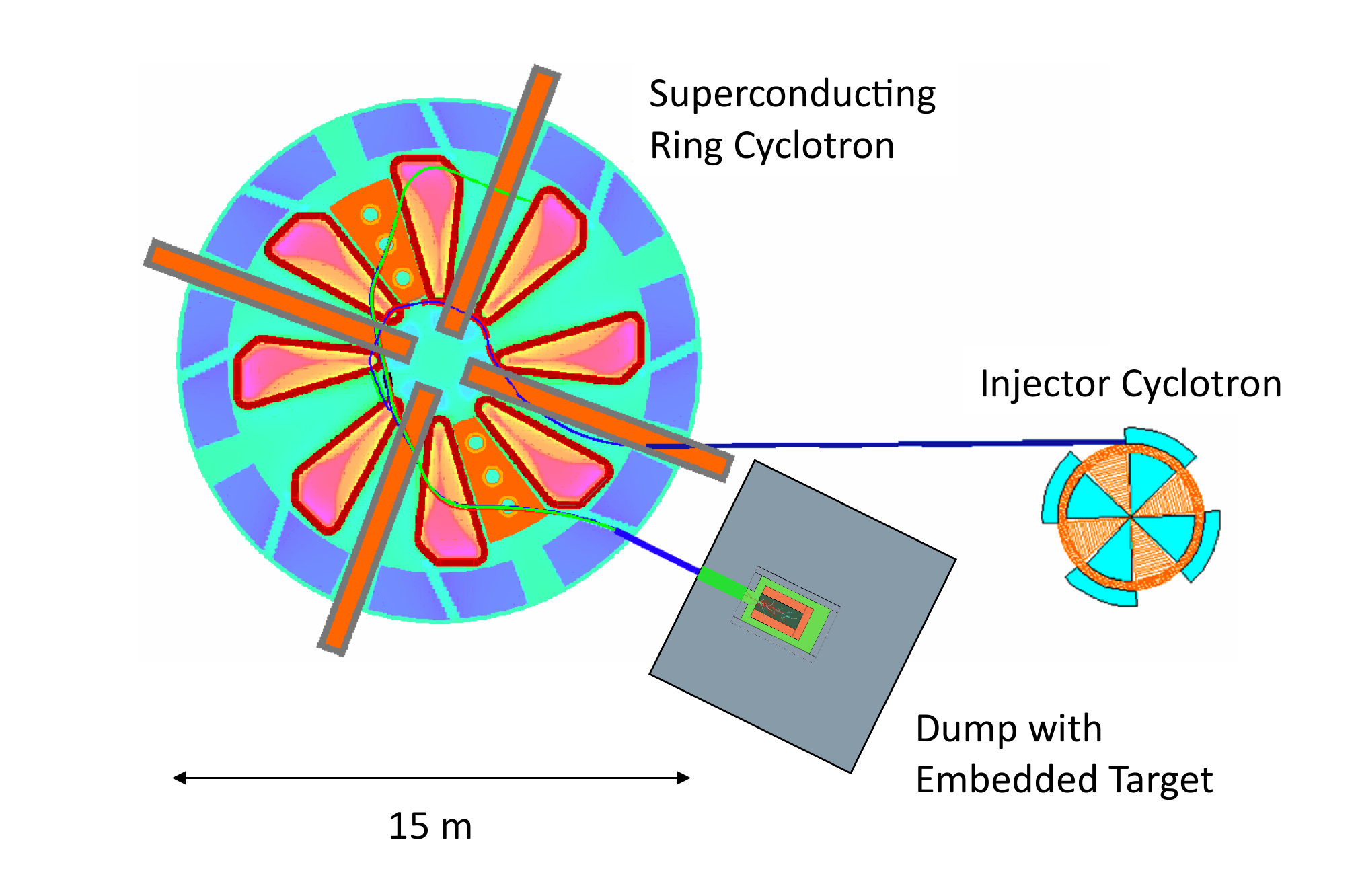}}}
\caption{\it \footnotesize  Layout of the two-cyclotron system.  The
  magnetic field regions are indicated by the pastel colors.  The
  injector cyclotron (small) is a compact design and the
  superconducting cyclotron (large) is a separated sector design.
  Figure from Ref.~\cite{pontecorvo}.
\label{cyclolayout} }
\end{wrapfigure}

Fig.~\ref{cyclolayout} shows the layout of the DAE$\delta$AUS two-cyclotron system.   The magnetic fields are indicated by
the pastel colors, where blue is negative and red is positive field.
The position of the RF in the valleys is indicated in the orange on
the DSRC.

The DSRC shares characteristics of 
several cyclotrons.
TRIUMF, in Vancouver, Canada, is a 500 MeV H$^-$ machine, over 18 meters in diameter 
(of beam orbit) but with a 
very low peak field to prevent Lorentz stripping of the very loosely-bound H$^-$ ions.  
The PSI cyclotron is 590 MeV, 12 m in diameter, and is
an 8-sector normal-conducting Ring Cyclotron.  PSI has the distinction of being the 
 world's highest power cyclotron, at 1.3 MW currently.  The RIKEN Superconducting 
 Ring Cyclotron in Wako, Japan, \cite{RIKEN} is designed for low-current heavy ion beams, but its 
 magnet shares almost identical engineering specifications to the requirements for the 
 DSRC.  This machine serves as a valuable engineering and costing model for the DSRC.

In Sec.~\ref{CP},  we broke the approach toward achieving the final
system into four phases that had accompanying physics goals:  1) ion
source, low energy
beam transport and inflection; 2) acceleration in the injector
cyclotron, and subsequent extraction;   3) acceleration in the
superconducting ring cyclotron,
and 4) producing the DAR fluxes (targeting and shielding).
We will refer to these phases in the more detailed information we
provide in the following sections.

\subsection{The Cyclotron Design Parameters \label{parameters} }

In this section we discuss the current design parameters of the
two cyclotrons:  the DAE$\delta$ALUS injector cyclotron (DIC) and the DAE$\delta$ALUS
Superconducting Ring Cyclotron (DSRC).    Progress on the design
on both machines has been published \cite{NIM, Yang, Minervini}.
This section records the design parameters of the leading options
for these two cyclotrons, at present.  We note that the designs are
under development and alternatives are under study.

The DIC can be used as a stand-alone accelerator for production of
isotope-decay-at rest fluxes, as in the IsoDAR sterile
neutrino search (Sec.~\ref{disapp}), as well as an injector for the
DAE$\delta$ALUS CP violation search.    The primary design option for the 
DIC is an H$_2^+$ accelerator.   The
design parameters are shown in Table~\ref{tab:injtab}.
\begin{table}[t]
\centering
\begin{tabular}{lrrllrrl}
\hline
$E_{max}$ & 	60 MeV/amu	&	$E_{inj}$ & 35 keV/amu \\
$R_{ext}$ &	1.99 m		&		$R_{inj}$ &55 mm  \\
$<B>$ @ $R_{ext}$ &1.16 T	 &	$<B>$ @ $R_{inj}$ &	0.97 T  \\
Sectors		& 4			&		Hill width	&	28 - 40 deg \\
Valley gap	& 1800 mm	& Pole gap	& 100 mm  \\
Outer Diameter & 6.2 m	 & Full height & 2.7 m  \\
Cavities	& 4					& Cavity type	& $\lambda/2$, double-gap  \\
Harmonic &	 6th		&			RF-frequency	& 49.2 MHz  \\
Acc. Voltage	& 70 - 250 kV	 & Power/cavity &	$<110$ kW  \\
$\Delta E$/turn	 &1.3 MeV	& Turns &107  \\
$\Delta R$ /turn @ $R_{ext}$	& $20$ mm	 & $\Delta R$/turn @ $R_{inj}$ & $>56$ mm  \\
Coil size & 200x250 mm$^2$ & Current density	 & 3.1 A/mm$^2$  \\
Iron weight & 450 tons	& Vacuum  & $< 10^{-7}$ mbar  \\
\hline
\end{tabular}
\caption{\it \footnotesize  DIC design parameters, as of Snowmass 2013.  \label{tab:injtab}}
\end{table} 

For IsoDAR, an important question under study is whether to use H$_2^+$, as it is planned for DAE$\delta$ALUS or instead run 
deuterons which, like H$_2^+$, has charge-to-mass $(q/m)$ ratio of $0.5$.      This represents a change in philosophy at the
target of IsoDAR, where the H$_2^+$ design requires neutron production
from the beryllium target, while a deuteron design delivers neutrons
directly to the target.  In this latter configuration, lower
beam energies can produce the required rate of
$^8$Li.    Thus the DIC central
region would be maintained, but we would employ a smaller accelerating
magnet radius for the lower energy beam.  This reduces costs and makes
underground construction easier although more beamline shielding will be
required.   During the period of the
Snowmass study, we made great progress on this alternative design.  We
will compare this to the H$_2^+$-based DIC design for IsoDAR at a review at the 
Eloisatron Workshop on High power cyclotrons and targets for Neutrino
Physics, Nov. 12-17, 2013.

\begin{table}[t]
\centering
\begin{tabular}{lrrllrrl}
\hline
$E_{max}$ & 	800 MeV/amu	&	$E_{inj}$ & 60 MeV/amu \\
$R_{ext}$ &	4.9 m		&		$R_{inj}$ & 1.8 cm  \\
$<B>$ @ $R_{ext}$ &1.88 T	 &	$<B>$ @ $R_{inj}$ &	1.06 T  \\
Sectors		& 8			&		Hill width
& 23  deg \\
$B_{max}$	& 6.05 T	& Pole gap	& 60 mm  \\
Outer Diameter & 14 m	 & Full height & 5.6 m  \\
Cavities	& 6					& Cavity type	& Single gap (4) \\
& & & Double gap (2)  \\
Harmonic &	 6th		&			RF-frequency	& 49.2 MHz  \\
Acc. Voltage	& 550-1000 kV	 & Power/cavity &	300 kW  \\
$\Delta E$/turn	 & 3.6  MeV	& Turns & 420  \\
$\Delta R$ /turn @ $R_{ext}$	& 5 mm	 & $\Delta R$/turn @ $R_{inj}$ & $>10$ mm  \\
\hline
\end{tabular}
\caption{\it \footnotesize  DSRC design parameters, as of Snowmass 2013. \label{tab:srctab}}
\end{table}

The DSRC is used to accelerate the 60 MeV/amu H$_2^+$ to 800 MeV/amu \cite{NIM} for 
production of pion/muon decay at rest fluxes \cite{firstpaper}.  The  
parameters of the DSRC are given in Table~\ref{tab:srctab}.
 Aspects of the 8-sector design are very similar to the RIKEN cyclotron
\cite{RIKEN}.  In Table~\ref{RIKsim}, 
we provide a comparison of parameters.  The primary difference is that 
RIKEN is a 6-sector machine.
Based on the success of the RIKEN accelerator and the engineering
complexities of an 8-sector design, the DSRC
is now proceeding through a second iteration where we are examining the
viability of a six-sector configuration.  We have completed an engineering
study of the new six-sector layout, which is published in Ref.~\cite{Minervini}.
This report 
develops a viable engineering design satisfying requirements for the superconductor, as well as
structural and cryogenic requirements. The work includes solid modeling
and analyses for the conductor 
and winding pack design, high
temperature superconductor and copper current leads for the magnet, structural design of
the magnet cold mass, cryostat and warm-to-cold supports, cryogenic design of the magnet
cooling system, and magnet power supply sizing.  A beam simulation
study for the six-sector design is now under way.

\begin{table}[t]
\centering
\begin{tabular}{lrrl}
\hline
\textbf{Basic Parameters} & \textbf{DSRC } & \textbf{RIKEN-SRC} & \textbf{Unit} \\
\hline
Maximum field on the hill& 6.05 &3.8 &T\\
Maximum field on the coil& 6.18 &4.2 &T\\
Stored Energy & 280 & 235 & MJ \\
Coil size & 30$\times$ 24 or 15$\times$ 48 & 21$\times$ 28 & cm$^2$ \\
Coil Circumference & 9.8 & 10.86 & m \\
Magnetomotive force& 4.9 & 4 & MAtot/sector \\
Current density & 34 & 34 & A/mm$^2$ \\
Height& 5.6 & 6.0 & m \\
Length& 6.9 & 7.2 & m \\
Weight& $\leqq$450 & 800 & ton \\
Additional magnetic shield& 0 & 3000 & ton/total \\
\hline
\textbf{Magnetic Forces} &&&\\
\hline
Expansion& 1.87 or 1.8 &2.6 &MN/m\\
Vertical& 3.7& 3.3 &MN \\
Radial shifting& 2.7& 0.36 &MN \\
Azimuthal shifting & 0.2 & 0 &MN \\
\hline
\textbf{Main Coil} &&&\\
\hline
Operational current& 5000&5000 &A\\
Layer $\times$ turn & 31$\times$16& 22$\times$18 & \\
Cooling& Bath cooling & Bath cooling & \\
Maddock Stabilized Current& 6345& 6665 &A \\
\hline
\textbf{Other Components} &&&\\
\hline
SC trim& no&4 &sets\\
NC trim $\times$ turn & no & 22 &pairs \\
Stray field in the SRC valley region& 0.7 & 0.04 & T \\
Gap for thermal insulation& 40 & 90@min. &mm \\
Extraction method& Stripper foil&Electrostatic channel& \\
\hline
\end{tabular}
\caption{\it \footnotesize Comparison of the DSRC design at the time
  of Smowmass 2013 (an 8-sector design) with RIKEN (a 6-sector
  design). \label{RIKsim}}
\end{table}

\subsection{Status of Present and Near Future Accelerator R\&D \label{nowandsoon}}

We have begun an extensive experimental testing program of the designs.
Our first tests are being performed in the spring/summer of 2013 at the Vancouver site of Best Cyclotron 
Systems, Inc.  The VIS, or Versatile Ion Source, a off-resonant
microwave discharge ion source  \cite{Maimone:2008zz} built at the Laboratori Nazionali del Sud (LNS) in Catania, Italy, has been shipped to Vancouver for these tests.  This 
source is designed to provide $\sim$50 mA of protons or deuterons and
is expected to reach $>$20 mA of H$_2^+$ when optimized. Best Cyclotrons has built a small test cyclotron 
suitable for 1 MeV (to ensure no neutron production), which allows 
captured beam to be accelerated to eight turns.
The central region,  consisting of a spiral inflector, dee
configuration, and post-collimator,  is being assembled as the
final Snowmass meeting begins.

\begin{wrapfigure}{l}{0.5\textwidth}
{\includegraphics[width=3.in]{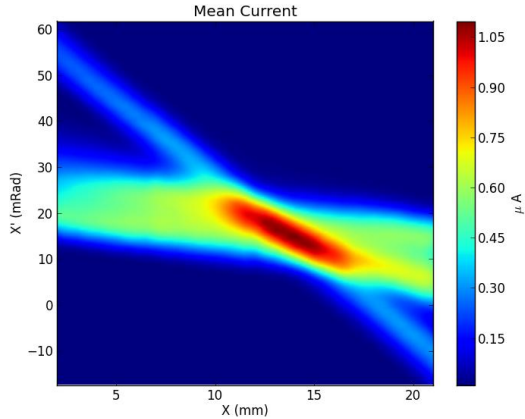}\hfill}
\caption{\it \footnotesize  
Emittance measurement result performed at the Best teststand.  Three
ellipses related to protons, H$_2^+$ and H$_3^+$ produced from the ion
source can be seen.
\label{emit}}
\end{wrapfigure}

Source characterization measurements and optimization for H$_2^+$ ions
have been started; the emittance plot shown in Fig.~\ref{emit}  is an
example.  Emittance is a measure of beam quality, plotted on $(x,x^\prime)$
or $(y,y^\prime)$ axes.  A good beam will be represented by an
ellipse.  The area of the beam ellipse is the emittance, which does
not change for conservative systems. 
In the plot shown in Fig.~\ref{emit}, one can discern three ellipses associated
with the three beam species:  protons (most tilted and weakest), H$_2^+$,
and slightly closer to the axis a small admixture of H$_3^+$.  The
different tilt of each ellipse reflects different focusing strengths
for the three beams.  Total beam current for
this run was $\sim10$ mA.
The ratio of the three species can be varied by changing the microwave
power fed to the source.  
This figure represents an optimum for H$_2^+$, with this ion comprising roughly 40\% of the beam.

 The measurements obtained for the run at Best Cyclotrons will provide
constraints for the DAE$\delta$ALUS OPAL simulations. 
The OPAL code  \cite{OPAL}, developed at the PSI, is widely used for simulations of highly space-charge-dominated beams in cyclotrons and beam-lines.  
Benchmarking code predictions with experimental results is a key element of the overall R\&D program.

While we are making substantial progress towards our ultimate goal with the tests in Vancouver, 
development of the complex central region often requires iterations, and we envision the need to improve the design 
based on these measurements.   As a result, we are already planning
the follow-up experiment: a new test-stand to be built at INFN-LNS in
Catania, Italy.  This will use an improved ion source, a redesigned transport line, and a 7 MeV/amu cyclotron.  The central 
 region of this new cyclotron will be designed based on the experience gained from the tests in Vancouver.  The magnetic field study 
 for the Catania teststand cyclotron has been completed, and Best Theratronics \cite{BEST}
 will provide the engineering study in 2013. The order for all cyclotron components (magnet, RF cavities, vacuum 
 system, diagnostic system, and electrostatic deflector) will be
 completed during winter 2013-14.   The timescale to construct 
 the components is one year.   Thus assembly would begin in late 2014,
 contingent on funding.

\subsection{Areas of R\&D -- A Summary for Each Phase \label{phases}}

During the Snowmass
Study, we were requested by the 
Frontier Capabilities Working Groups, as well as the Facilities
Panel for DOE High Energy Physics, to discuss the R\&D needs for each
phase.   This section briefly reviews the information we provided.

Phase
I is well underway.  As discussed above, the INFN-Catania ion source 
is now running in the teststand at Best Cyclotron Systems, Inc. 
Simulating the inflector is quite difficult, and
so development of this piece, and the overall central region, requires
experimental iterations.    Our second iteration on the test stand,
also described above, will be performed at Catania.

Beyond this, an interesting technical challenge for the $CP$ program, 
which runs H$_2^+$ at 800 MeV/amu,  is the removal of ions in the high vibrational
states.     At 800 MeV, the high vibrational 
states of this molecule will Lorentz-strip
in the 6 T outer field of the DSRC.  As much as 10\% of the beam could
be lost through this process if the ions are not removed from the beam.   Calculations show that the
lowest four states will be stable \cite{Haxton}.     We are
investigating ways to remove the high vibrational states.  Work in
collaboration with Oak Ridge National Laboratory retested the methods of Sen, {\it et al.} \cite{Sen},
which involve introducing a noble gas into the ion source.
Results were difficult to interpret in the first round of tests;
however,  dissociation of these vibrational states by this mechanism
in the ion source 
is believed to require long (millisecond) residency times of the ions
prior to extraction. 
If this is true, then we must consider a redesign of the
source \cite{Sen} or a method of removing the vibrational states after the
H$_2^+$ exits the source.

Phase II is also underway.   The overal goal isl output of 5 mA of
H$_2^+$ (or 10 mA of protons) at 60 MeV/amu.  The generalized perveance, which characterizes the
space charge of the cyclotron, is, at injection, equivalent to
existing machines.  An important paper
relevant to the space-charge issues of this cyclotron is Ref. \cite{Yang}.  Note that this is a full
OPAL simulation,  which is the well-established and experimentally benchmarked code
for these type of studies (it is the GEANT-equivalent for cyclotrons).
Our first iteration of this cyclotron passed an internal
review held at Erice in November 2012, and a second iteration of the
design parameters has already been established.       We plan another
review of the Phase I and Phase II combined results for next October, to
be held at PSI.

The primary issue for the Phase III development is 
the DRSC.  After a 
careful review, we have moved from the eight-sector design of
Ref.~\cite{NIM} to a 6-sector design which is very close to the
existing RIKEN cyclotron.   Simulations of the new machine are
underway.   
As discussed above, we have had a detailed engineering
study performed of this six-sector design \cite{Minervini}.  This has allowed a detailed
US-style costing for the magnet, which is the most expensive novel
element within DAE$\delta$ALUS.   Other costs can be determined
directly from the RIKEN and PSI experience 
with the correction for the differences in accounting.  

The 1 MW target is part of the Phase III development.  Because the beam is
extracted via stripping foils, we can extract to several targets to
limit instantaneous power.  Also, the beam will be painted over a 30 cm
target face, greatly reducing power issues.  

Phase IV, which is at the least advanced stage, takes the Phase III
system to high power.   The collaboration has several competing
conceptual designs on how to achieve this. The ideas presently under 
consideration will be carefully reviewed in the
future, as this final phase develops.

To briefly summarize the top experimental
issues on which the collaboration is working at present:  
\begin{enumerate}
\item   Increasing current of the source to $>$50 mA H$_2^+$. 
\item    Demonstrating the removal of high vibrational states of H$_2^+$. 
\item   Demonstrating 5 mA beam capture and emittance control in the central region 
	of the injector cyclotron.
\item Detailed simulation of high-efficiency beam extraction from the injector. 
\item Full end-to-end simulation of beam dynamics using the proven OPAL code.
\end{enumerate}
A long-term experimental goal is likely to involve a full-scale
prototype of one sector of the DSRC.  We are in the process of planning
for this.

The collaboration working on this R\&D includes universities,
international laboratories, and industry.  Along with members of Best
Cyclotrons, Inc., we have also collaborated with scientists from other
cyclotron companies (IBA \cite{IBA}, and AIMA \cite{AIMA}) on aspects
of our design.

In summary, we are proposing a step-wise approach for the development
of the components of the 
DAE$\delta$ALUS cyclotrons.    No single component must be pushed 
orders of magnitude beyond existing technology.   But when these smaller
steps 
are combined, the system represents a substantial leap forward.

\section{Broader Impacts of These Machines}

 {\it  Accelerators for America's Future} has stated: ``The
  United States, which has traditionally led the development and
  application of accelerator technology, now lags behind other nations
  in many cases, and the gap is growing.  To achieve the potential of
  particle accelerators to address national challenges will require
  sustained focus on developing transformative technological
  opportunities...'' \cite{AAF}.
Cyclotrons are a clear example.  Despite being originally developed 
in the US,  most cyclotron research and companies are
located outside of the US.  The major laboratories involved in
this initiative (INFN-Catania, PSI, and RIKEN) are outside the US.  On
the other hand, the universities involved in this program are largely
US-based.  
This allows for technology transfer and ensures the
next generation of cyclotron physicists in the U.S.  Through this, the program
serves a valuable national interest.

This evolving R\&D program provides examples of synergy 
between the goals of fundamental physics research and the needs of
society.   This has motivated close collaboration between
laboratories, universities, and industry on this project.   
In a time when R\&D funding for physics is decreasing, 
this program illustrates a cost-effective approach for development of tools for basic science.  

As a very immediate example, the Catania test-stand is drawing
substantial support from the private sector.  Best Medical Italy is
actively contributing to the collaboration.  Underlining the value of
the test-stand,  Krishnan Suthanthiran, president of TeamBest (parent
company of the Best family of companies) wrote in a letter of support,
``[The] original motivation for the device is for it to become the
injector for a very high intensity neutrino source for pure science
research (DAE$\delta$ALUS). The same concepts you have described have
an immediate medical radioisotope application.''  Our project
represents an excellent example of the societal value of basic
accelerator research.  The value to medical isotope production arises
from two aspects of the test-stand.  The first is development of high-current proton beams which ultimately enhance the production rate of
isotopes.    The second is that the system being developed can
accelerate any ion with the same charge-to-mass ratio as H$_2^+$,
including He$^{++}$ and deuterons.  In particular, $^{211}$At is
optimally produced by 28 MeV alpha (He$^{++}$) beams (note that 28 MeV
alphas have 7 MeV/amu, as per our machine design).  This isotope, being a pure alpha-emitter, is a
very powerful therapeutic agent.  Its
widespread clinical use is limited today only by the availability of
production capacity  \cite{At211}.  The test cyclotron developed in this project,
coupled with an existing commercial ion source for doubly stripped
helium ions \cite{pantechnik}, can immediately be applied to the production of this
isotope.  Alpha beams can also be applied to the production of
carrier-free $^{99}$Mo, and numerous other isotopes of commercial
interest.

\begin{table}[t]
\centering
\caption{\it \footnotesize Medical isotopes relevant to IsoDAR
  energies, from Ref. \cite{cost-cycl-2005}. }
{\footnotesize
\begin{tabular}{|l|c|c|}
\hline
Isotope & half-life & Use \\ \hline
$^{52}$Fe & 8.3 h &  The parent of the PET isotope $^{52}$Mn  \\
   & & and iron tracer
  for red-blood-cell formation and brain uptake studies.\\  \hline
$^{122}$Xe & 20.1 h &  The parent of  PET isotope $^{122}$I used to study
brain blood-flow. \\ \hline
$^{28}$Mg  & 21 h & A tracer that can be used for bone studies,
analogous to calcium \\ \hline
$^{128}$Ba  & 2.43 d & The parent of positron emitter $^{128}$Cs. \\
  & & As a
  potassium analog, this is used for heart and blood-flow imaging. \\ \hline
$^{97}$Ru & 2.79 d & A $\gamma$-emitter used for spinal fluid and liver
studies. \\ \hline
$^{117m}$Sn & 13.6 d & A $\gamma$-emitter potentially useful for bone
studies. \\ \hline
$^{82}$Sr & 25.4 d &  The parent of positron emitter $^{81}$Rb, a
  potassium analogue   \\ 
& &  This isotope is also directly used as a PET
  isotope for heart imaging. \\ 
\hline
\end{tabular}}
\label{tab:med}
\end{table} 

The injector cyclotron can be a powerful tool for isotope production.
As a source of 60 MeV protons at 600 kW, it represents  substantially
higher beam power  than available at existing isotope machines.
This could enable significantly greater yield.  It should be noted that
along with the improved cyclotron,  a new system of targeting must be
developed to handle the higher power.   
Time was devoted during this
Snowmass study to developing a concept to address this, and more
details are provided in Ref.~\cite{pontecorvo}, now in draft. 
Also, most existing machines are in the 30 to 40 MeV range.  Thus
the energy opens new opportunities also.
A list of isotopes that are
uniquely produced in the 60 to 70 MeV range is provided in 
Table~\ref{tab:med}.  The isotope which is especially valuable is 
$^{82}$Sr, the parent to $^{81}$Rb.
Finally, 
making use of the fact that ions of the
same charge-to-mass ratios can also be accelerated,
$\alpha$ beams can be produced.  These have currents limited
only by the availability of He$^{++}$ ion sources.  There is
tremendous potential associated with these beams.    
A document related to IsoDAR medical applications has been posted on the arxiv \cite{alonsomed}.

The DAE$\delta$ALUS SRC takes the next step 
beyond PSI,   increasing the extracted energy from 590
to 800 MeV,  with a factor of 5 in current.  This makes this machine 
a member of the GeV-scale, 10-MW-class of accelerators.   These are
sought-after machines for ``ADS''
(Accelerator-Driven Systems)---used for nuclear waste
transmutation, driving of sub-critical reactors (e.g. thorium), and
tritium production.   Many proposals exist, but 
cost has been a major impediment to their
realization.  To date, only one such project has progressed to the
advanced R\&D and construction phase: this is MYRRHA \cite{myrrha}
in Mol, Belgium.   

The DAE$\delta$ALUS cyclotron design began as as concept for ADS \cite{ADS}.
Our cyclotrons are attractive because they will have substantially
reduced cost over the linacs, which
until now have been viewed as the only viable technology to reach
these levels of beam power at the GeV energy range.  This could
facilitate deployment of more than one machine, maintaining production when any
particular accelerator is serviced.  With successful
development of these relatively inexpensive cyclotrons, a substantial growth in the ADS field
can be anticipated.    The application of DAE$\delta$ALUS machines to
ADS technology is under study by scientists at Brookhaven National
Laboratory \cite{BNL}.   Their design uses two full power accelerator modules
(DIC+DSRC) and one half-power.

\section{Conclusions}

This whitepaper has summarized the status and plans of the
DAE$\delta$ALUS program.  The cyclotrons open up a rich physics
program, where only a subset is presented here.    
Progress toward realizing these machines has been excellent and a clear R\&D program has been laid out.

It is apparent that there are substantial studies required to prove the
DAE$\delta$ALUS design.  We note that while the R\&D
aspects of each piece of the system are challenging, they are not
orders of magnitude beyond what has been accomplished.   The
remarkable step forward occurs when the smaller steps are combined. 
Thus, as the final Snowmass meeting begins,  the
DAE$\delta$ALUS program is well placed to advance neutrino physics as
well to bring cyclotron innovation back to its birthplace.

\end{document}